\documentclass[journal]{IEEEtran}
\usepackage{cite}
\ifCLASSINFOpdf
  \usepackage[pdftex]{graphicx}
  \graphicspath{{../pdf/}{../jpeg/}}
\DeclareGraphicsExtensions{.pdf,.jpeg,.png}
\else
\fi
\usepackage{makecell}
\usepackage{threeparttable}
\usepackage{amsmath}
\usepackage{algorithmic}
\usepackage{color}
\usepackage{array}

\usepackage{amsfonts,amssymb}
\usepackage{booktabs}
\usepackage{multirow}
\usepackage{multicol}
\begin{document}
\title{Exploring Resolution Fields for Scalable Image Compression with Uncertainty Guidance}

\author{Dongyi~Zhang,~Feng~Li,~Man~Liu,~Runmin~Cong,~\IEEEmembership{Senior~Member,~IEEE},~Huihui~Bai, ~Meng~Wang,~\IEEEmembership{Fellow,~IEEE}~and~Yao~Zhao,~\IEEEmembership{Fellow,~IEEE}
\thanks{
This work was supported in part by National Key R \& D Program of China (No.2022YFE0200300), the National Natural Science Foundation of China (No. 61972023,
62002014, 62120106009), the Beijing Natural Science Foundation (4222013, L223022), the Beijing Nova Program under Grant (Z201100006820016), and the CAAI-Huawei MindSpore
Open Fund. Corresponding author: Huihui Bai.

Dongyi Zhang, Man Liu, Huihui Bai, and Yao Zhao are with the Beijing Key Laboratory of Advanced Information Science and Network Technology, Beijing, 100044, China; and also with the Institute of Information Science, Beijing Jiaotong University, Beijing 100044, China. Email: \{dyzhang, manliu, hhbai, yzhao\}@bjtu.edu.cn.

Feng Li and Meng Wang are with the School of Computer Science and Engineering, Hefei University of Technology, Hefei, 230009, China. Email: fengli@hfut.edu.cn, eric.mengwang@gmail.com.

Runmin Cong is with the School of Control Science and Engineering, Shandong University, Shandong, 250100, China. Email: rmcong@sdu.edu.cn.}
}

\markboth{Journal of \LaTeX\ Class Files,~Vol.~14, No.~8, August~2015}%
{Shell \MakeLowercase{\textit{et al}.}: Bare Demo of IEEEtran.cls for IEEE Journals}
\maketitle
\begin{abstract}
Recently, there are significant advancements in learning-based image compression methods surpassing traditional coding standards. Most of them prioritize achieving the best rate-distortion performance for a particular compression rate, which limits their flexibility and adaptability in various applications with complex and varying constraints. In this work, we explore the potential of resolution fields in scalable image compression and propose the reciprocal pyramid network (RPN) that fulfills the need for more adaptable and versatile compression. Specifically, RPN 
first builds a compression pyramid and generates the resolution fields at different levels in a top-down manner. 
The key design lies in the cross-resolution context mining module between adjacent levels, which performs feature enriching and distillation to mine meaningful contextualized information and remove unnecessary redundancy, producing informative resolution fields as residual priors. 
The scalability is achieved by progressive bitstream reusing and resolution field incorporation varying at different levels. Furthermore, between adjacent compression levels, we explicitly quantify the aleatoric uncertainty from the bottom decoded representations and develop an uncertainty-guided loss to update the upper-level compression parameters, forming a reverse pyramid process that enforces the network to focus on the textured pixels with high variance for more reliable and accurate reconstruction. Combining resolution field exploration and uncertainty guidance in a pyramid manner, RPN can effectively achieve spatial and quality scalable image compression. Experiments show the superiority of RPN against existing classical and deep learning-based scalable codecs. Code will be available at \textcolor{red}{https://github.com/JGIroro/RPNSIC}.
\end{abstract}
\begin{IEEEkeywords}
Scalable image compression, reciprocal pyramid network, resolution field, cross-resolution context mining, uncertainty-guided loss.
\end{IEEEkeywords}
\IEEEpeerreviewmaketitle

\section{Introduction}
\IEEEPARstart{I}{mage} compression aims at reducing the redundancy in an image while preserving its visual quality, which is a fundamental and critical technique in computer vision. 
With the growing popularity of diversified image-collecting devices and image circulation in social media, the volume of image data communicated over networks has increased dramatically, which leads to higher demands for efficient and multifunctional image coding.
To achieve faster transmission and less storage of visual data, various image compression algorithms have been developed and made significant advances in the past few decades.

The most common image compression methods~\cite{watchcases1992jpeg,skodras2001jpeg,bellard2017bpg,balle2016end,balle2018variational,balle2017end,lee2018context,minnen2018joint,cao2022end,limu,zhu2022unified,cheng2020learned} follow the typical ``transform--quantization--entropy'' coding pipeline.
These methods can be divided into traditional image compression and learning-based image compression according to whether end-to-end deep learning technology is adopted. 
Although the former traditional methods~\cite{watchcases1992jpeg,skodras2001jpeg,bellard2017bpg} adjust the compression rate by changing the quantization coefficient, which lacks global optimization capability as each module is manually designed separately.
In recent years, Learning-based methods~\cite{balle2016end,balle2018variational,balle2017end,lee2018context,minnen2018joint,cao2022end,cheng2020learned,limu,zhu2022unified} have shown stronger performance benefiting from end-to-end optimization. 
However, these approaches typically train a model for a specific point on the rate-distortion (R-D) curve. This means multiple compressed bitrates require multiple models, which is resource-consuming and inconvenient due to device storage or network transmission constraints in various multimedia applications.

To fulfill the diversified need of various bitrates in practical applications, some variable-rate approaches~\cite{toderici2015variable,cai2018efficient,johnston2018improved,choi2019variable} enable different compression rates within only one trained model. However, they overlook the potential of leveraging the scalability of bitstreams to reduce the redundancy of information in network transmission. Some works~\cite{icip2020,dpict,schwarz2007overview,boyce2015overview,bai2021learning,mei2021learning} propose scalable image compression, which allows generating a bitstream by once coding and decoding appropriate subsets of the whole bitstream to reconstruct complete images with multiple compression levels.
Lee~\emph{et al.}~\cite{dpict} represent the latent tensor in ternary digits and present a trit-plane-based quality-scalable codec.
Bai~\textit{et al.}~\cite{bai2021learning} propose to control the quantization step of residual coefficient between the predicted and accurate reconstruction for quality scalability.
Nevertheless, spatial scalability from encoded bitstreams is also important for information transmission in resource-restricted conditions.
Mei~\emph{et al.}~\cite{mei2021learning} take both spatial and quality scalability into consideration and construct a multi-layer architecture for learning-based scalable image compression.
However, this method lacks consideration for contextualized information across different resolutions as it simply uses stacked convolutions and upscaling modules to incorporate predicted latent representations from the previous compression layer.

Actually, the high-resolution features contain abundant information that is conducive to detail recovery and redundant information that needs to be filtered.
The low-resolution features well summarize the global context of images but miss lots of details in textured regions.
It is necessary to mine the informative contexts at different resolutions for more effective compression.
Besides, all the methods mentioned above pursue deterministic predictions without considering the uncertainty of neural networks~\cite{gawlikowski2021survey}. 
In image generation, such uncertainty reveals the instability of reconstructed pixels, which is high-related to the fidelity of textures and edges recovery.

To tackle these problems, we explore the potential of resolution fields and propose the reciprocal pyramid network (RPN) with uncertainty guidance to achieve both spatial and quality scalable image compression. 
First, we convert the original image to various resolution versions, formulating a multi-level forward pyramid process that produces informative resolution fields and performs top-down compression. 
Between two adjacent levels, we design a cross-resolution context mining module (CRCM) composed of alternate information enhancement blocks (IEB) and redundancy removal blocks (RRB) to learn contextualized information from the previous level (lower resolution).
The IEB aims at aggregating the features of all positions at the higher-resolution space to enrich the contextualized information of low-resolution feature representations.
The RRB investigates the intrinsic channel and spatial sparsity of enriched features at low resolution for redundancy removal and important context distillation.
With iteratively IEBs and RRBs in CRCM, we can learn the resolution field from the previous level as the residual prior, which is flowed into the current level to help higher-resolution compression.
Hence, by the progressive low-to-high resolution compression of RPN, the bitstream can support different rates, thereby ensuring spatial and quality scalability.

Then, we construct a reverse bottom-up pyramid process to progressive model the aleatoric uncertainty of reconstructed images. 
Different from the forward top-down pyramid, between two levels, we characterize the textured pixels with large variance from higher resolution as the uncertainty parts.
The predicted uncertainty map is utilized as guidance to re-update the compression parameters of the lower-resolution result with uncertainty-guided loss.
Therefore, our RPN can achieve more reliable and accurate reconstruction, especially for low-rate/resolution compression levels.

The contributions of this work are summarized as follows:

\begin{itemize}
\item
We propose a reciprocal pyramid network (RPN) that explores the resolution fields for spatial and quality scalable image compression with uncertainty guidance. Experiments demonstrate the superiority of RPN against existing state-of-the-art methods.

\item We propose the cross-resolution context mining module (CRCM) that performs feature enriching and distillation to mine meaningful contextualized information and remove unnecessary redundancy, producing informative resolution fields as residual priors for reducing bitrates without quality degradation.

\item We design an uncertainty-based optimization scheme that enforces RPN to focus on the textured pixels with high variance by using uncertainty-guided loss in a high-to-low manner, which can achieve better R-D performance without extra computational cost during testing.

\end{itemize}

The remainder of this paper is organized as follows. We first discuss related work in Section II. Then we introduce our proposed method for scalable image compression in section III. The experimental results are presented in Section IV, and Section V concludes this paper.

\section{RELATED WORK}

\subsection{Learning-Based Image Compression}
Deep learning in image compression has made significant progress in recent years.
Ball{\'e}~\emph{et al.}~\cite{balle2016end} propose the pioneering
introduce convolutional neural network (CNN) into image compression, which optimizes the R-D performance end-to-end.
The authors further design a hyperprior network~\cite{balle2018variational} to capture the spatial dependencies of latent representations as side information for the entropy model, which becomes a typical paradigm widely used in later works~\cite{lee2018context,minnen2018joint,cao2022end,limu}.
Minnen \emph{et al.}~\cite{minnen2018joint} generalize the hierarchical Gaussian scale mixture (GSM)~\cite{wainwright1999scale} model to a Gaussian mixture model (GMM) and combine it with autoregressive model for better R-D performance.
Cao~\textit{et al.}~\cite{cao2022end} predict the posterior distributions of bottleneck representation with deep Gaussian process regression to achieve a novel 2-D context-based entropy model.
Zhu~\textit{et al.}~\cite{zhu2022unified} propose a vectorized prior-based unified multivariate Gaussian mixture to speed up compression procedure.
Cheng~\textit{et al.}~\cite{cheng2020learned} leverage discretized
Gaussian mixture likelihoods and attention modules to improve the network capability with moderate training complexity.
Tang~\textit{et al.}~\cite{tang2022joint} introduce graph attention to explore the long-range relationships into high-level abstract representations for learned image compression.
Inspired by the success of transformer architecture in computer vision tasks~\cite{vaswani2017attention,liu2021swin}, Zou~\textit{et al.}~\cite{zou2022devil} propose a symmetrical transformer framework with absolute transformer blocks in the encoder and decoder, which achieves the state-of-the-art.
There are also some methods~\cite{paml,extreme2019,highfidelity2020} that exploit the powerful generation ability of generative adversarial network (GAN) to balance the perceptual and R-D performance.

\begin{figure*}[!t]
\centering
\includegraphics[width=1.0\linewidth]{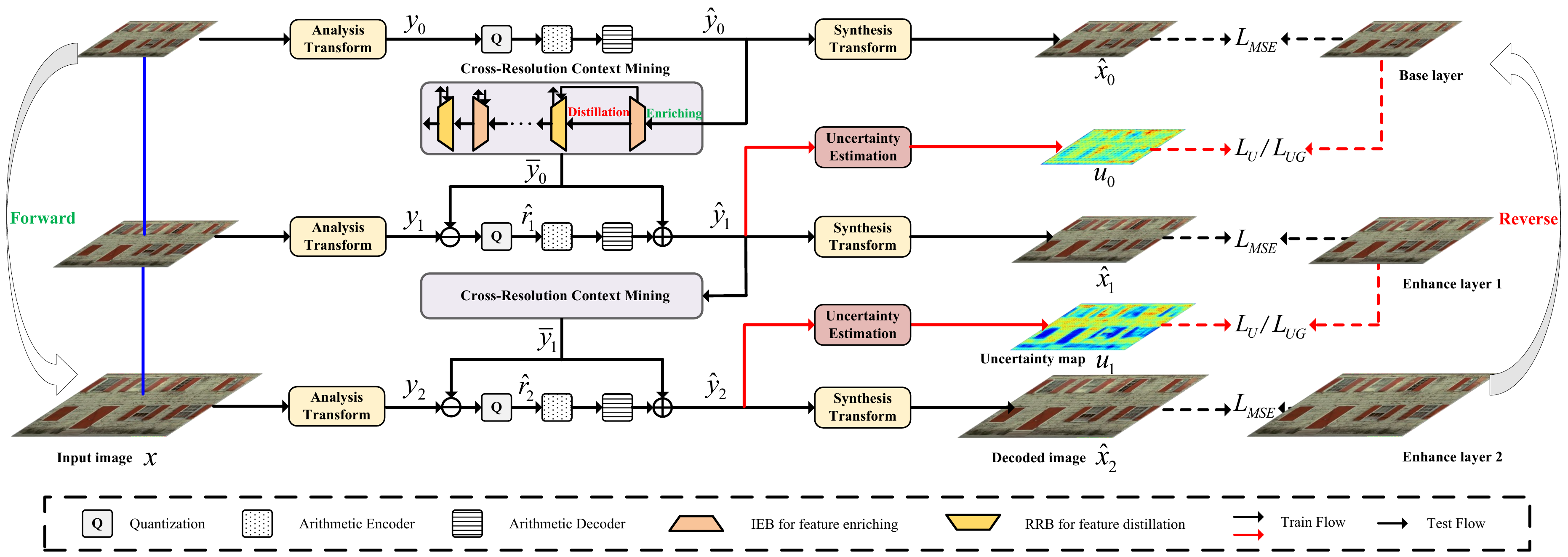}
\caption{Overview of the proposed RPN for spatial scalability. In the forward pyramid process, analysis transforms convert the original image to various-resolution versions. 
With the cross-resolution context mining module, only residuals between the layers need to be transmitted.
In the reverse bottom-up pyramid process, the uncertainty map estimated with high variance from the bottom serves as the guidance to re-update the upper compression parameters with uncertainty-guided loss. }
\label{fig_1}
\end{figure*}

\subsection{Scalable Image Compression}
The definition of scalable coding was first developed in the extension of the traditional coding standard H.264/AVC~\cite{schwarz2007overview}, termed SVC, which uses a multi-layer structure to produce decoded images/videos with different resolutions or quality. 
Later, High Efficiency Video Coding (HEVC)~\cite{sullivan2012overviewHEVC} standard further proposes the
Scalable High-efficiency Video Coding (SHVC)~\cite{boyce2015overview} that additionally introduces signal-to-noise ratio and color gamut scalability functionalities than prior SVC. 

Recently, deep learning also demonstrates its outstanding effectiveness in the scalable image compression task.
Early methods~\cite{toderici2015variable,toderici2017full,johnston2018improved} utilize recurrent neural network (RNN)-based codecs to perform progressive compression, which achieves coarse granular scalability.
Bai~\textit{et al.}~\cite{bai2021learning} design a quality scalable auto-encoder scheme by controlling the quantization step of residual coefficient between the predicted and accurate reconstruction. 
Yang~\emph{et al.}~\cite{slimmable} propose a slimmable compressive autoencoder with a rate and complexity control mechanism, which can be easily adapted to obtain scalable bitstreams.
Ma~\emph{et al.}~\cite{ma2022deepfgs} propose a learned fine-grained scalable image compression network which provides a fine-grained flexible bitstream covering a full bitrate range.
Lee~\emph{et al.}~\cite{dpict} propose a trit-plane algorithm to tackle the progressive image compression problem, in which the estimated bitstream can be truncated at any point to support fine granular scalability.
Mei~\textit{et al.}~\cite{mei2021learning} learn to explore the potential for feature domain prediction and reuse between layers to realize image compression at multiple resolutions and qualities.
Besides these human perception-oriented approaches, some researchers transfer to study scalable compression methods~\cite{sun2020semantic,choi2022scalable,zhang2023rethinking} toward machine perception.
Sun~\emph{et al.}~\cite{sun2020semantic} present a semantically structured image coding network to generate semantically structured bitstreams, where each part of the bitstream denotes a particular item and is capable of analysis or reconstruction.
Choi~\emph{et al.}~\cite{choi2022scalable} present a 3-layer scalable codec to support the transition of latent image representation from simpler to more complex tasks for scalability in both human and machine vision.
Zhang~\emph{et al.}~\cite{zhang2023rethinking} propose a scalable cross-modality compression method that forms different granularity representations and reconstructs visual information by scalable coding based on semantic, structural, and signal layers.

\subsection{Uncertainty in Deep Learning}
The uncertainty in deep neural networks can be classified into aleatoric uncertainty and epistemic uncertainty, which are related to the noise inherent in collected data and the ignorance of the model corresponding to data respectively~\cite{kendall2017uncertainties}.
Franchi~\textit{et al.}~\cite{franchi2022latent} propose an effective deterministic uncertainty method to better analyze features from unseen images in light of the knowledge acquired by CNN.
Zheng~\textit{et al.}~\cite{zheng2022pointras} refine the predictions under the uncertainty selection criterion to remedy the potential information deficiency of lower-resolution point clouds.
By using the Bayesian neural network,~\cite{dusenberry2020efficient} and~\cite{wilson2020bayesian} estimate uncertainty with the posterior distribution of the model weights to marginalize the likelihood distribution at inference for image classification. 
Ning~\textit{et al.}~\cite{ning2021uncertainty} introduce image super-resolution (SR) into a Bayesian estimation framework to model the uncertainty (variance) in SR results with an adaptive weighted loss.
As far as we know, there is few work that has investigated the uncertainty in image compression. In this paper, we present RPN to simultaneously reconstruct images while modeling the aleatoric uncertainty by incorporating an uncertainty estimation module.

\section{PROPOSED METHODS}

\subsection{Overview}

The overall architecture of the proposed reciprocal pyramid network (RPN) is illustrated in~Fig.~\ref{fig_1}, using a (4-2-1) structure for spatial scalability as an example. 
Notation (4-2-1) indicates the resolution of compressed images from the $4\times$ downsampled version to the original one, forming a 3-level forward top-down pyramid compression process (left in Fig.~\ref{fig_1}).
Our network is inspired by existing learning-based compression codecs~\cite{johnston2018improved,mei2021learning} that utilize the decoded feature representations from previous level to help current level compression. 
Instead of simply stacking multiple stages with recurrent connections or shallow convolutional layers to bridge the bitstreams between every two adjacent levels, we incorporate a cross-resolution context mining module (CRCM) that operates on previous features before passing to the current level.
CRCM iteratively consists of information enhancement block (IEB) and redundancy removal block (RRB) for feature enriching and distillation, which can effectively mine the contextualized information to produce informative resolution fields as the residual priors participating in the current level compression. 
For quality scalability, the downsampling operations are removed.
In RPN, the lowest-resolution/quality level is denoted as the base layer which is responsible for the lowest spatial/quality compression. The other two are denoted as enhance layers that perform compression on larger-resolution/quality images. Each level in RPN is a subnetwork composed of analysis transform, quantization, arithmetic coding, and synthesis transform.

\begin{table}[t]
    \renewcommand\arraystretch{1.2}
  \centering
  \caption{The Parameter Setting of analysis transform and synthesis transform. ``Conv'' and ``DConv'' denote convolutional and deconvolutional layer, respectively. ``IGDN'' is the inverse operation of ``GDN''. $C_l$ is the channel number corresponding to layer $l$. $\downarrow2$ and $\uparrow2$ denote $2\times$ feature downsampling and upsampling.}
        \begin{tabular}{c|c}
        \hline
        \hline
        \textbf{Analysis Transform} & \textbf{Synthesis Transform} \\
        \hline
        
        Conv + GDN & DConv + IGDN \\
        ($5\times 5 \times C_l$, $\downarrow2$) & ($5 \times 5 \times C_l$, $\uparrow2$)\\
        \hline
        Conv + GDN & DConv + IGDN \\
        ($5\times 5 \times C_l$, $\downarrow2$) & ($5 \times 5 \times C_l$, $\uparrow2$)\\
        \hline
        Conv + GDN & DConv + IGDN\\
        ($5\times 5 \times C_l$, $\downarrow2$) & ($5 \times 5 \times C_l$, $\uparrow2$)\\
        \hline
        Conv & DConv\\
        ($5\times 5 \times C_l$, $\downarrow2$) & ($5 \times 5 \times C_l$, $\uparrow2$) \\
        \hline
        \hline
        \end{tabular}%
  
  \label{tab1}%
\end{table}   %

\textbf{\emph{Compression at Base Layer}}. We adopt the architecture in~\cite{balle2018variational} as the backbone, where the detailed parameter setting is shown in Table~\ref{tab1}. Let $x$ denote the original RGB image with the size of $H\times W\times 3$, where $H$ and $W$ are the height and width of $x$, respectively. As for the ``4-2-1'' three-level model, the input for the base layer is the $4\times$ downsampled image denoted as $x_0$ with the size of $H/4\times W/4\times 3$. At the encoder side, we use an analysis transform module composed of 4 convolutional layers with generalized divisive normalization (GDN) activation to obtain the latent compact representation $y_0$. Then, quantization followed by an arithmetic encoder (AE) is adopted to generate the bitstream of $y_0$. After an arithmetic decoder (AD), we can obtain a quantized representation $\hat{y}_0$ that serves as the input fed into the synthesis transform module and CRCM. The synthesis transform module adopts a symmetric structure as the analysis transform module, which performs feature upsampling and reconstructs a compressed image $\hat{x}_0$ with the same size as $x_0$. The whole process of the base layer compression $\mathcal{B}_0$ can be formulated as 
\begin{equation}
\label{eq4}
    \begin{aligned}
    \hat{x}_0 &= \mathcal{B}_0({x_0}) \\
     &= \mathcal{S}_0(Q(\mathcal{A}_0(x_0))),
    \end{aligned}
\end{equation}
where $\mathcal{S}_0(\cdot)$ and $\mathcal{A}_0(\cdot)$ represent the synthesis transform and analysis transform operations, respectively. $Q(\cdot)$ is a quantization function~\cite{balle2017end} that adds uniform noise during training while round $y_0$ to its nearest integer during inference.

\textbf{\emph{Compression at Enhance Layer}}. For two adjacent compression levels $l-1$ and $l$, the quantized representation from the $(l-1)$-th level is denoted as $\hat{y}_{l-1}$. At the current level, we employ the same subnetwork as the base layer to transform the corresponding input image $x_l$ to a latent representation $y_l$. Besides, we leverage a CRCM to generate a resolution field $\bar{y}_{l-1}$ from $\hat{y}_{l-1}$, which can be expressed as 
\begin{equation}
\label{eq5}
\bar{y}_{l-1} = CRCM_{l-1}(\hat{y}_{l-1}),
\end{equation}
where $CRCM_{l-1}(\cdot)$ denotes the CRCM function operates on $\hat{y}_{l-1}$. Different from the base layer, we only quantize the residue between $y_l$ and $\bar{y}_{l-1}$ for AE and AD. Besides, to fully exploit the resolution field for better reconstruction, $\bar{y}_{l-1}$ is also used as the auxiliary information combined with the decoded representation for synthesis transforming. Therefore, we can reconstruct the $l$-th level compressed image with the enhance compression $\mathcal{E}_l$ as 
\begin{equation}
\label{eq6}
\begin{aligned}
\hat{y}_{l} &= \mathcal{E}_l(x_l, \bar{y}_{l-1}) \\
&= \mathcal{S}_l(Q(\mathcal{A}_l(x_l)-\bar{y}_{l-1}) + \bar{y}_{l-1}).
\end{aligned}
\end{equation}
where $\mathcal{S}_l(\cdot)$ and $\mathcal{A}_l(\cdot)$ represent the synthesis transform and analysis transform operations of $l$-th level, respectively. As we can see, the resolution field is not only beneficial to the encoder side that enables residue-based coding and transmission, but also supports high-quality reconstruction at the decoder side without additional bitrate. 

In this way, our RPN can provide scalable bitstreams for various versions of given images through a progressive top-down compression manner. Besides, by incorporating the bitstreams from the lower into higher layers, RPN can gradually improve the rate-distortion performance, achieving spatial and quality scalability.

\textbf{\emph{R-D Optimization}}. At any compression level $l~(0 \leq l < L)$, for the corresponding image $x_l$, we use the entropy model~\cite{rissanen1981universal} to estimate its rate $R_l$ of quantized latent representation $\hat{y}_l$, which can be formulated as
\begin{equation}
\label{eq1}
R_l = \left\{
\begin{array}{lcr}  
    -\mathbb{E}_{y_l}[\log_2p(\hat{y}_l)]& \mbox{for}\; l=0 \\
     -\mathbb{E}_{r_l}[\log_2p(\hat{r}_l)] + \sum\limits_{0<t<l}R_t + R_0 & \mbox{for}\; l\neq 0 \\ \end{array} \right.,
\end{equation}
where $\hat{r}_l$ is the quantized residual representation at level $l$ and $p(\cdot)$ denotes the probability density function. 
$t$ represents the intermediate level number between the base layer ($l=0$) and the $l$-th layer in RPN.
We use the MSE between compressed image $\hat{x}_l$ and its original input image $x_l$ to measure the distortion. Thus, the loss of each level can be defined as
\begin{equation}
\label{eq2}
\begin{split}
\mathcal{L}_l &= D_l + \lambda_l R_l,
\\ D_l &= MSE(x_l,\hat{x}_l),
\end{split}
\end{equation}
where $\lambda_l$ controls the trade-off between rate $R_l$ and distortion $D_l$. Consequently, the total loss $\mathcal{L}_{sca}$ for the forward scalable compression in RPN is 
\begin{equation}
\label{eq3}
\mathcal{L}_{sca} = \sum^{L-1}_{l=0}\mathcal{L}_l.
\end{equation}

After the first R-D optimization step, we additionally construct a reverse bottom-up pyramid process that progressively conducts explicitly aleatoric uncertainty estimation from high-to-low resolutions. Similarly, between two levels, we characterize the texture
pixels with high variance from bottom level as the uncertainty parts. Then we estimate the uncertainty map that serves as the guidance to re-update upper level compression parameters with uncertainty-guided loss, enabling RPN to make more reliable and accurate reconstructions, especially for low-resolution/quality stages.

\begin{figure}[t]
\centering
\includegraphics[width=1\linewidth]{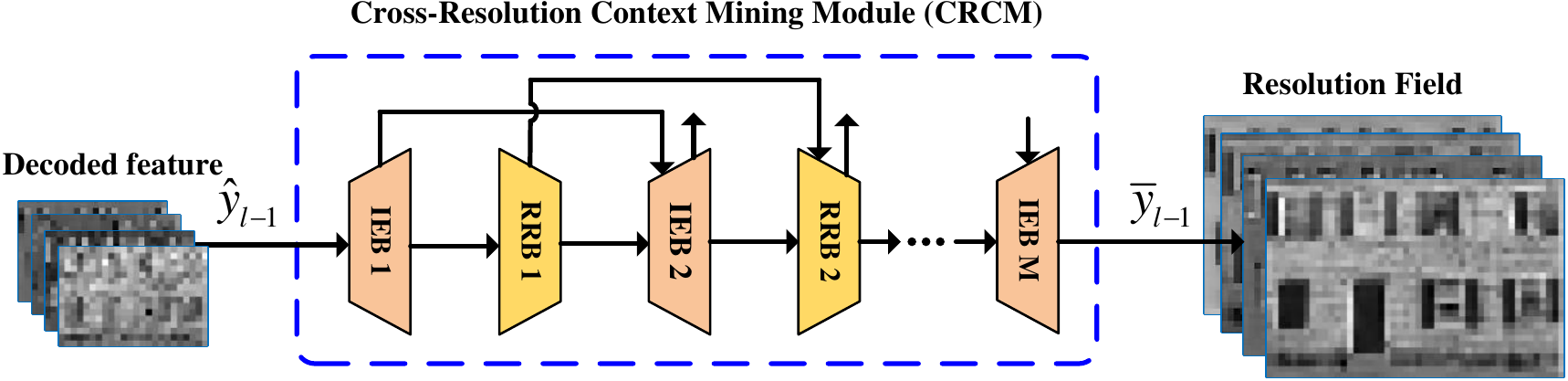}
\caption{Structure of the cross-resolution context mining module (CRCM) which consists of the iterative information enhancement blocks (IEB) and redundancy removal blocks (RRB).}
\label{fig_2}
\end{figure}

\subsection{Cross-Resolution Context Mining Module}
In RPN, the cross-resolution context mining module (CRCM) is introduced between two adjacent compression levels to learn informative resolution fields for better compression quality. The structure of CRCM is shown in Fig.~\ref{fig_2}. It consists of information enhancement block (IEB) and redundancy removal block (RRB) iteratively to mine the contextualized information from two perspectives. First, IEB receives the features from previous states and applies a global context attention mechanism to enrich the context attending in higher-resolution space. Second, with the enriched feature, RRB aims to investigate its intrinsic channel and spatial sparsity to remove the unimportant redundancy and distill more important context that is conducive to detail recovery.
For spatial scalability, supposing there are $M$ IEBs and $N$ RRBs in one CRCM, due to the iterative implementation, $N$ is just one more than $M$, \emph{i.e.} $N=M+1$.
As for quality scalability, $N$ is equal to $M$.

\begin{figure}[t]
\centering
\includegraphics[width=1.0\linewidth]{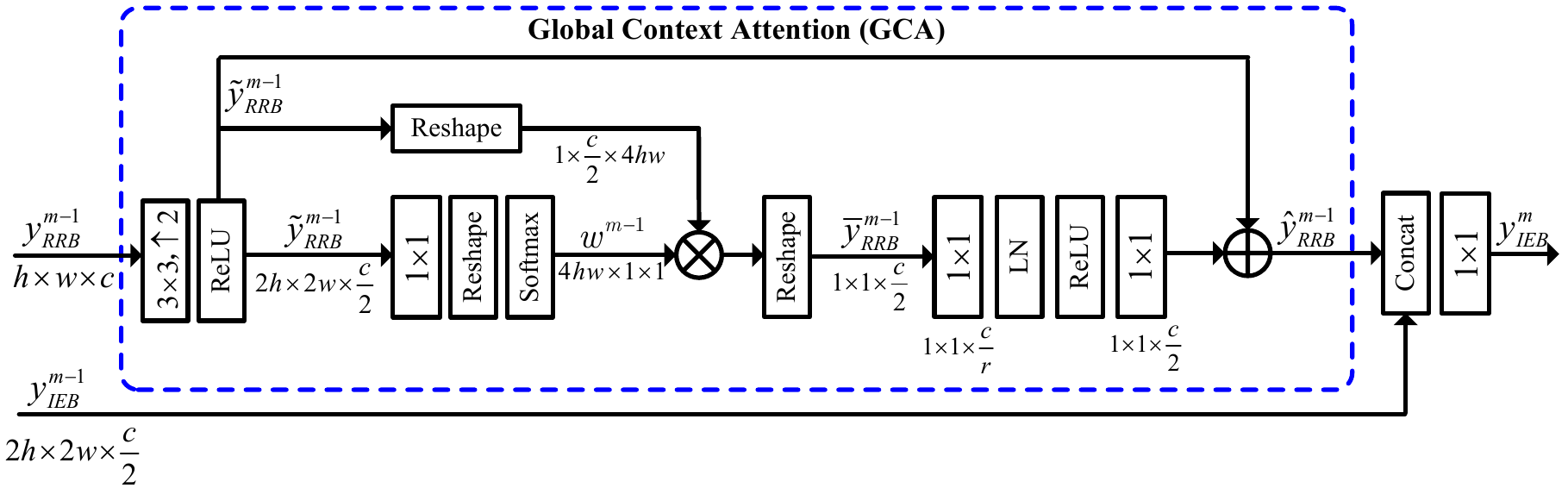}
\caption{Structure of the information enhancement block (IEB) with global context attention (GCA) mechanism.}
\label{fig_3}
\end{figure}

\textbf{\emph{Information Enhancement Block}}.
 For the $m$-th IEB, the incoming features $y^{m-1}_{IEB}\in\mathbb{R}^{2h\times 2w\times \frac{c}{2}}$ and $y^{m-1}_{RRB}\in\mathbb{R}^{h\times w\times c}$ are from the $(m-1)$-th IEB and the $(m-1)$-th RRB, respectively. Here, $h$, $w$, and $c$ denote the height, width, and channels.
The detailed structure of IEB is shown in Fig.~\ref{fig_3}. 
In the $m$-th IEB, we build a global context attention (GCA) mechanism that aggregates the features of all positions at the higher-resolution space to form a global feature with more contextualized information. To be specific, a $3\times 3$ deconvolution with stride 2 is first used to upsample $y^{m-1}_{RRB}$ to a magnified version $\tilde{y}^{m-1}_{RRB}\in\mathbb{R}^{2h\times 2w\times \frac{c}{2}}$. Then, we employ a $1\times 1$ convolution with reshape and softmax operations to compute the attention weight $w^{m-1}\in\mathbb{R}^{4hw\times 1 \times  1}$. Hence, by weighted averaging on $\tilde{y}^{m-1}_{RRB}$ with $w^{m-1}$ among all query positions followed by a reshape operation, we can obtain the global attentive context feature $\bar{y}^{m-1}_{RRB}\in\mathbb{R}^{1 \times 1 \times \frac{c}{2}}$. This process can be formulated as
\begin{equation}
\label{eq7}
    \bar{y}^{m-1}_{RRB} = \sum^{N_p}_{j=1}w^{m-1}_{j}\tilde{y}^{m-1}_{RRB,j},
\end{equation}
where $N_p$ denotes the total query positions in $\tilde{y}^{m-1}_{RRB}$ which is reshaped to the size of $1 \times \frac{c}{2} \times 4hw$ for multiplication. Motivated by~\cite{cao2020global}, two $1\times 1$ convolutional layers with the channel reduction ratio $r$ are used  as the bottleneck transformation module to capture the channel-wise interdependencies of $\bar{y}^{m-1}_{RRB}$ and produce the aggregated context feature $\hat{y}^{m-1}_{RRB}$ with element-wise addition that
\begin{equation}
    \label{eq8}
    \hat{y}^{m-1}_{RRB} = \tilde{y}^{m-1}_{RRB} + \mathbf{w}_2*ReLU(LN(\mathbf{w}_1*\bar{y}^{m-1}_{RRB})),
\end{equation}
where $LN(\cdot)$ and $ReLU(\cdot)$ denote the layer normalization (LN) and ReLU activation in Fig.~\ref{fig_3}. $\mathbf{w}_{1}\in\mathbb{R}^{1\times 1\times \frac{c}{r}}$ and $\mathbf{w}_{2}\in\mathbb{R}^{1\times 1\times \frac{c}{2}}$ are the paramter sets of two $1\times 1$ convolution with $\frac{c}{r}$ and $\frac{c}{2}$ filters, respectively. Finally, $\hat{y}^{m-1}_{RRB}$ is in combination with the incoming feature $y^{m-1}_{IEB}$ to generate the enriched feature $y^{m}_{IEB}\in\mathbb{R}^{2h\times 2w \times \frac{c}{2}}$ of the $m$-th IEB by
\begin{equation}
    \label{eq9}
    y^{m}_{IEB} = \mathbf{w}_3*Concat(\hat{y}^{m-1}_{RRB}, y^{m-1}_{IEB}),
\end{equation}
where $Concat(\cdot)$ represents feature concatenation and $\mathbf{w}_3$ is the parameter set of the final $1\times 1$ convolution for feature integration.

\textbf{\emph{Redundancy Removal Block}}. On the one hand, feature enrichment with IEB can improve the amount of information and facilitate better detail expression, but on the other hand, it will also bring redundant information, such as structural contexts, that affects compression efficiency. To alleviate this problem, we investigate the intrinsic sparsity of features and propose the RRB to condense redundant information and distill more important contexts. Existing researches~\cite{gumbel,dynamic,wang2021exploring} have demonstrated the effectiveness and efficiency of spatial or channel sparsity for vision tasks. In this work, we also explore the sparsity for compression from spatial and channel perspectives. 

\begin{figure}[t]
\centering
\includegraphics[width=1.0\linewidth]{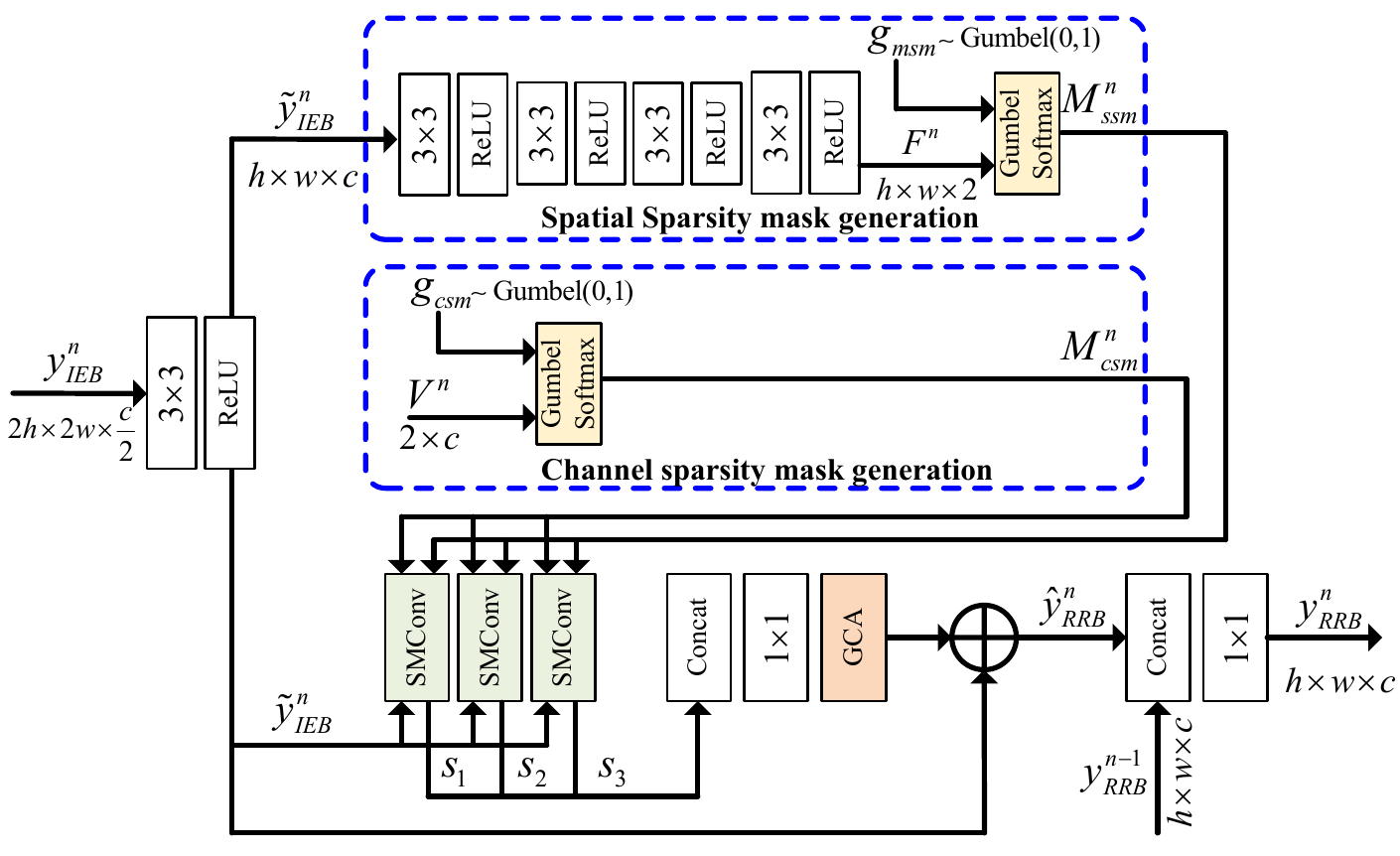}
\caption{Structure of the redundancy removal block (RRB) which exploits channel and spatial sparsity for redundancy removal and feature distillation.}
\label{fig_4}
\end{figure}

Let $y^{n}_{IEB}\in\mathbb{R}^{2h\times 2w\times \frac{c}{2}}$ and $y^{n-1}_{RRB}\in\mathbb{R}^{h\times w\times c}$ be the output of the $n$-th IEB and the $(n-1)$-th RRB. As shown in Fig.~\ref{fig_4}, in the $n$-th RRB, we first resample $y^{n}_{IEB}$ to its lower-resolution version by a $3\times 3$ convolution with stride 2 and model the spatial and channel sparsity based on Gumbel-Softmax distribution~\cite{gumbel}. Specifically, similar to~\cite{wang2021exploring}, for the channel sparsity, we first initialize a two-category variable $V^n\in\mathbb{R}^{2\times c}$ with random values. And then we feed it into a Gumbel softmax layer to generate a binary channel sparse mask $M^{n}_{csm}\in\mathbb{R}^c$ that measures whether the channels are informative or not (redundant), where ``1'' denotes the positive to be preserved and ``0'' is aborted. With two samples (denoted as $g_{csm}[i], i=1, 2$) drawn from Gumbel(0, 1) distribution, we can obtain the channel-wise mask by the softmax function as
\begin{equation}
    \label{eq10}
    M^{n}_{csm}[c] = \frac{\exp(V^n[1, c] + g_{csm}[1, c])/\tau}{\sum^2_{i=1}\exp(V^n[i,c]+g[i, c])/\tau},
\end{equation}
where $\tau$ is a temperature parameter.

For spatial sparsity, we also use the Gumbel softmax distribution-based strategy to learn a binary spatial sparse mask that identifies whether the regions in observed features are informative. Therefore, as shown in Fig.~\ref{fig_4}, for the resampled feature $\tilde{y}^{n}_{IEB}\in\mathbb{R}^{h\times w\times c}$, we can obtain the spatial-wise mask $M^{n}_{ssm}$ by
\begin{equation}
    \label{eq11}
    M^{n}_{ssm}[h, w] = \frac{\exp(F^n[1, h, w] + g_{msm}[1, h, w])/\tau}{\sum^2_{i=1}\exp(F^n[i,h, w]+g[i, h, w])/\tau},
\end{equation}
where $F^n[1, h, w]\in\mathbb{R}^{h\times w\times 2}$ is produced by a four-layer submodule. $g_{msm}$ also denotes the samples drawn from Gumbel$(0, 1)$ distribution.

Note that the purpose of RRB is just to utilize the sparsity for redundancy removal and feature distillation, here we directly leverage the sparse mask convolution (SMConv) in~\cite{wang2021exploring} as the backbone without specific design that uses these sparse masks as guidance to learn more informative features. Next, a GCA module (see Fig.~\ref{fig_3}) followed by a shortcut is further applied to learn a global attentive context feature as follow:
\begin{equation}
    \label{eq12}
    \begin{aligned}
    &s_k = SMConv_k(M^{n}_{csm}, M^{n}_{msm}, \tilde{y}^{n}_{IEB})  \quad (k=1, 2, 3),\\
    &\hat{y}^{n}_{IEB}=\tilde{y}^{n}_{IEB} + GCA(\mathbf{w}_4 * Concat(s_1, s_2, s_3)),
    \end{aligned}
\end{equation}
where $GCA(\cdot)$ denotes the function of GCA. $SMConv_k(\cdot)$ denotes the function of the $k$-th SMConv. $\mathbf{w}_4\in\mathbb{R}^{c\times1\times1}$ is the weight set of $1\times1$ convolution to integrate $s_1$, $s_2$, and $s_3$. Finally, by fusing $\hat{y}^{n}_{IEB}$ and the feature $y^{n-1}_{RRB}$ from the $(n-1)$-th RRB via feature concatenation and a $1\times 1$ convolution, the output $y^{n}_{RRB}$ of the $n$-th RRB can be obtained by 
\begin{equation}
    \label{eq13}
    y^{n}_{RRB}= \mathbf{w}_5 * Concat(\hat{y}^{n}_{IEB}, y^{n-1}_{RRB}),
\end{equation}
where $\mathbf{w}_5$ denotes the parameter set of the final $1\times 1$ convolution.

In summary, our CRCM can achieve alternative feature enriching and distillation by implementing IEBs and RRBs iteratively, thus gradually mining the contexts from both high- and low-resolution spaces, finally producing informative resolution fields as the residual prior to help the next compression level (see Fig.~\ref{fig_2}). 

\begin{figure*}[t]
\centering
\setlength{\abovecaptionskip}{-0.2cm}
\includegraphics[width=1\linewidth]{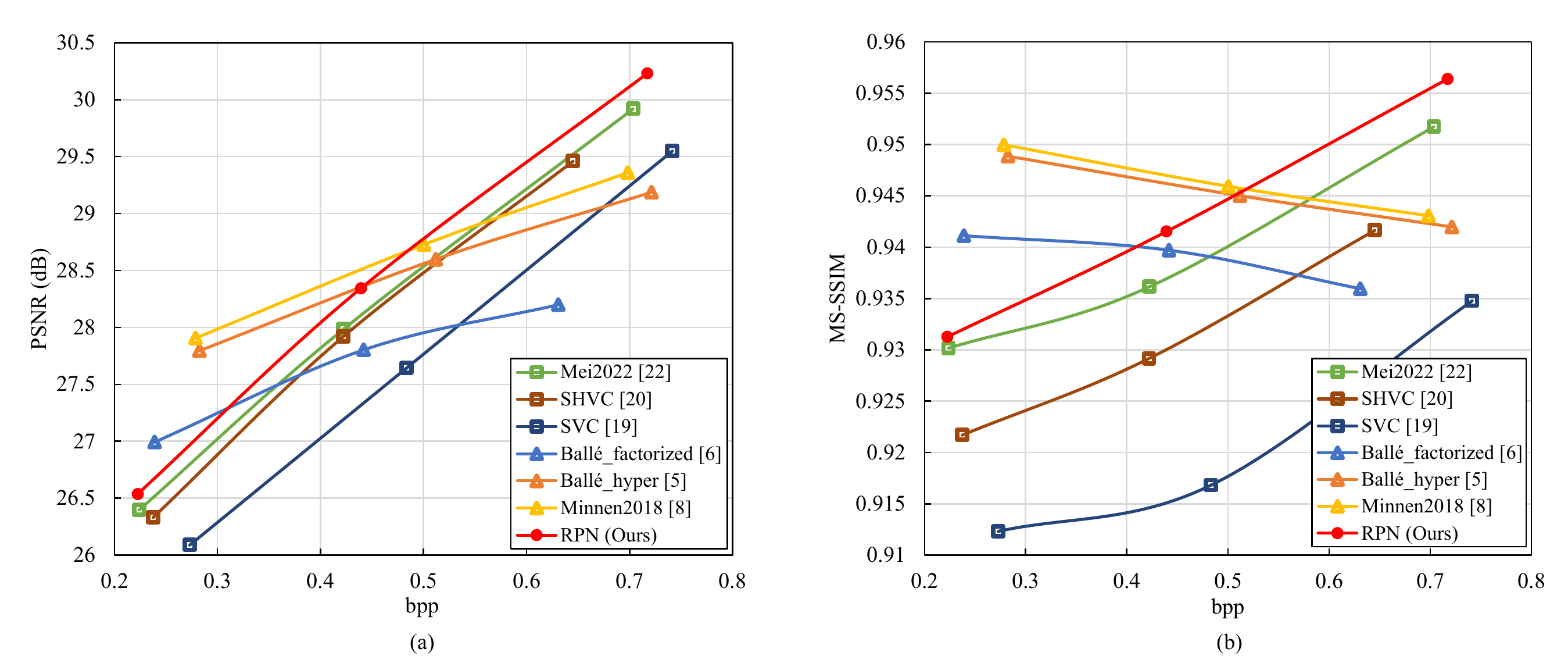}
\caption{Rate-distortion performance evaluation with existing compression methods at spatial scalability in terms of (a) PSNR and (b) MS-SSIM.}
\label{fig5}
\end{figure*}
\subsection{Uncertainty-Guided Reverse Pyramid Optimization}
The goal of image compression is to reconstruct a new one with lower storage while preserving the image details. Actually, due to the quantization process, lots of texture and edge information are lost, which makes it hard to recover a vivid image, especially on low spatial resolution or quality. Therefore, in RPN, as shown in Fig.~\ref{fig_1}, we construct a reverse bottom-up pyramid process that progressively conducts explicitly aleatoric uncertainty estimation from high-to-low resolutions. Specifically, let $g_{l}(\cdot)$ denote the codec of the $l$-th compression level and $\hat{x}_l$ be the reconstructed image of $x_l$. Given the decoded representation $\hat{y}_{l+1}$ from the $(l+1)$-th level, We first insert an uncertainty estimation module $f^e_{l}(\cdot)$ to learn the uncertainty $\delta_{l}$ by characterizing the texture pixels with high variance in $\hat{x}_{l}$ via a probabilistic manner. Thus, the coding model of the $l$-th level can be reformulated as 
\begin{equation}
\label{eq14}
\begin{aligned}
    \hat{x}_l &= g_{l}(x_l) = x_l + \epsilon \delta_{l} \\
    &=x_l + \epsilon f^e_{l}(\hat{y}_{l+1}),
\end{aligned}
\end{equation}
where $\epsilon$ represents the Gaussian distribution with zero-mean and unit-variance, which is assumed for characterizing the likelihood function by
\begin{equation}\label{eq15}
p(\hat{x}_l | x_l,\delta_l) = \frac{1}{\sqrt{2\pi\delta_l}}\exp(-\frac{\left| \left| \hat{x}_l-x_l \right| \right|_2}{2\delta_l}).
\end{equation}
Here, $\hat{x}_l$ and $\delta_l$ also denote the mean (\emph{i.e.} reconstructed image) and the variance (\emph{i.e.} uncertainty). Motivated by~\cite{ning2021uncertainty}, we also impose Jeffrey's prior~\cite{figueiredo2001adaptive} $p(w) \propto \frac{1}{w}$ on $\delta_l$. Therefore, Eq.~\ref{eq15} can be reformulated as 
\begin{equation}\label{eq16}
\begin{split}
p(\hat{x}_l,\delta_l | x_l) & \propto p(\hat{x}_l | x_l,\delta_l)p(\delta_l) \\
&= \frac{1}{\sqrt{2\pi\delta_l}}\exp(-\frac{\left| \left| \hat{x}_l-x_l \right| \right|_2}{2\delta_l})\frac{1}{\delta_l} \\
&=\frac{1}{\sqrt{2\pi}\delta_l^{\frac{3}{2}}}\exp(-\frac{\left| \left| \hat{x}_l-x_l \right| \right|_2}{2\delta_l}).
\end{split}
\end{equation}
Then, a log-likelihood formulation is leveraged to further rephrase $\delta_l$
\begin{equation}
\label{eq17}
\ln p(\hat{x}_l | x_l,\delta_l) = -\frac{\left| \left| \hat{x}_l-x_l \right| \right|_2}{2\delta_l}-\frac{3}{2}\ln\delta_l - \frac{1}{2}\ln2\pi.
\end{equation}
We use a Gaussian likelihood to model the aleatoric uncertainty, which will lead to a minimization objective on given $x_l$
\begin{equation}
\label{eq18}
\mathcal{L}_U(\theta_{l}) = \frac{1}{D}\exp(-\ln2\delta_l)\left| \left| \hat{x}_l-x_l \right| \right|_2 + \frac{3}{2}\ln\delta_l,
\end{equation}
where the former residual regression term is associated with the reconstruction quality that re-updates the parameter $\theta_l$ of the subnetwork conditioned on the uncertainty. The latter term is used for uncertainty regularization. In this work, we denote $\ln\delta_l$ as the uncertainty map $u_l$ for the $l$-th level and set $D=1$ for our image-level compression task. To further reinforce our network to focus on optimizing the pixels with higher uncertainty, we introduce an uncertainty-guided loss $\mathcal{L}_{UG}$ that uses the uncertainty map $u_l$ as a scaling weight to improve the distortion loss (Eq.~(1))
\begin{equation}\label{eq19}
\mathcal{L}_{UG} =  u_l \left| \left| \hat{x}_l-x_l \right| \right|_2.
\end{equation}

According to Eq.~(\ref{eq14}), Eq.~(\ref{eq18}), and Eq.~(\ref{eq19}), the proposed RPN can model the uncertainty of the lower-resolution compressed image by observing the learned latent representations from its higher-resolution compression level, resulting in better textures recovery. With such bottom-up progress, we can progressively improve the compression effectiveness on low spatial/quality conditions.

\subsection{Training Strategy}
\label{strategy}
As the analysis above, in combination with R-D and uncertainty-guided optimization, the whole process of RPN contains three steps: 1) Singly forward R-D optimization using the scalability loss $\mathcal{L}_{sca}$ (Eq.~(1)) to ensure appropriate bitrate and distortion that enables uncertainty to be efficiently obtained without being disturbed; 2) Implementing $\mathcal{L}_{U}$ with $\mathcal{L}_{sca}$ to encourage the network to reversely estimate the uncertainty map via the uncertainty estimation module; 3) Combining $\mathcal{L}_{sca}$ and $\mathcal{L}_{UG}$ to guide the network to focus on unreliable pixels, thus further improving the reconstruction quality.

\section{EXPERIMENTAL RESULTS}

\subsection{Implementation Details}
\textbf{Dateset Setting and Evaluation Metrics.} Following~\cite{mei2021learning}, we use the CLIC~\cite{CLIC2020} dataset as our training data, which contains 1633 images. During the training, the original images are randomly cropped into $512 \times 512$ patches for spatial scalability and $256 \times 256$ patches for quality scalability.
The widely used Kodak~\cite{franzen1999kodak} dataset is used for performance comparison.
In addition, the same downsampling operation as in \cite{mei2021learning} is used to generate $2 \times$ and $4 \times$ downscaled images for spatial scalable implementation.
We evaluate the compression performance by calculating the average R-D performance in terms of peak signal to noise ratio (PSNR) and multi-scale structural similarity (MS-SSIM)~\cite{wang2004image}.

\textbf{Training setting.} For spatial scalability, the model is trained for $1200k$ iterations using Adam optimizer \cite{kingma2014adam} with the batch size of 8. The channel number $C_l$ of each level is set as 96.
The initial learning rate is set to $1e-4$ both in Step-1 and Step-2 for $200k$ iterations, and drops to $3e-5$ in Step-3 for the last $800k$ (see Section~\ref{strategy}).
For quality scalability, the model is trained for $2500k$ iterations with $600k$, $400k$, and $1500k$ rounds for the three stages, respectively. The channel number is set as 96, 120, 144 and 196 from the top layer to the bottom layer.
We conduct the experiments using the Tensorflow \cite{abadi2016tensorflow} and Tensorflow-compression \cite{balle2018tensorflow} libraries over one NVIDIA 3090Ti GPU with 24GB memory.

\begin{figure*}[t]

\centering
\setlength{\abovecaptionskip}{0.1cm}
\includegraphics[width=1\linewidth]{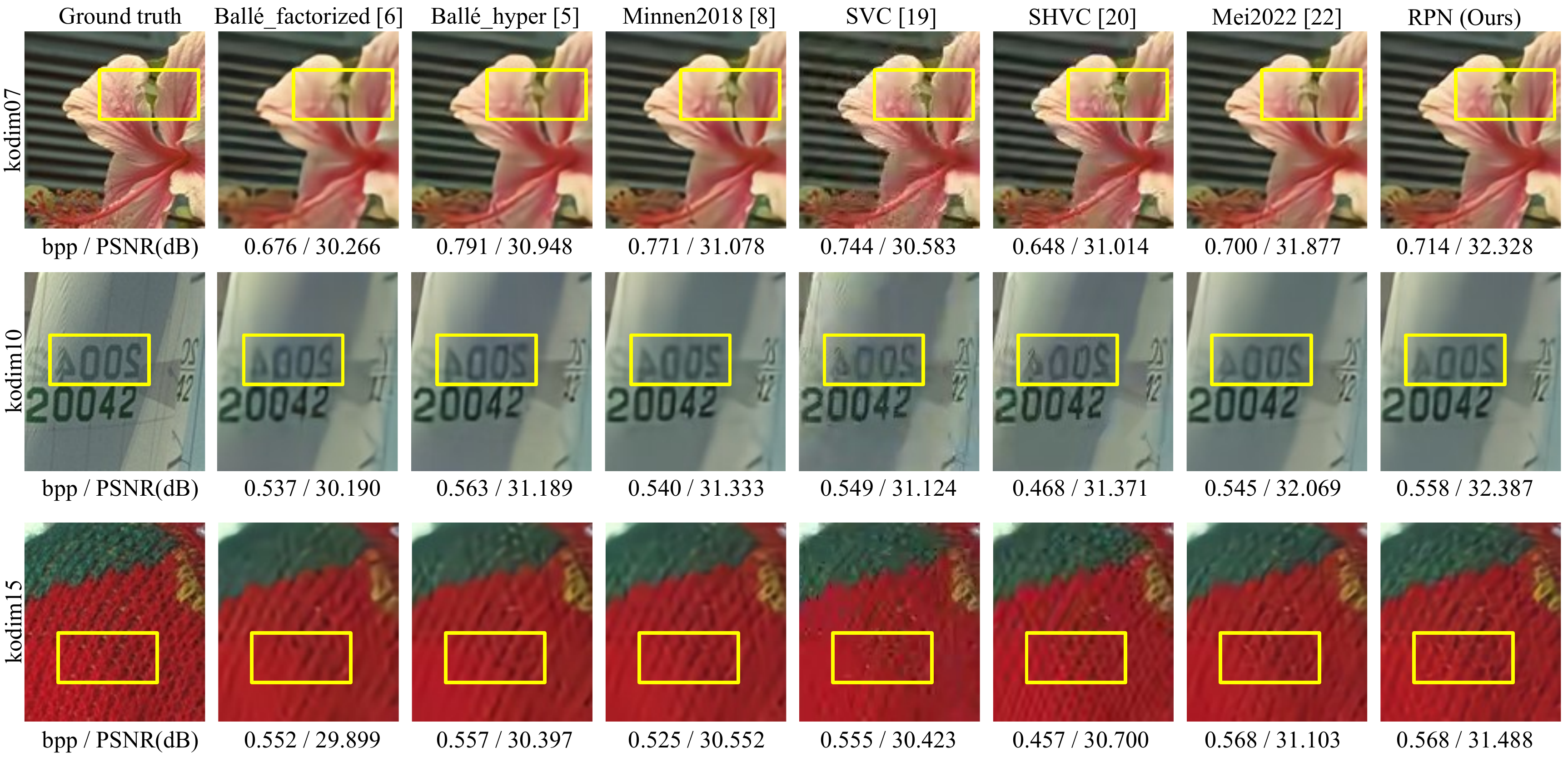}
\caption{Qualitative results of the final enhancement layer for different methods at spatial scalability.}
\label{fig6}

\vspace{0.2cm} 

\centering
\includegraphics[width=1\linewidth]{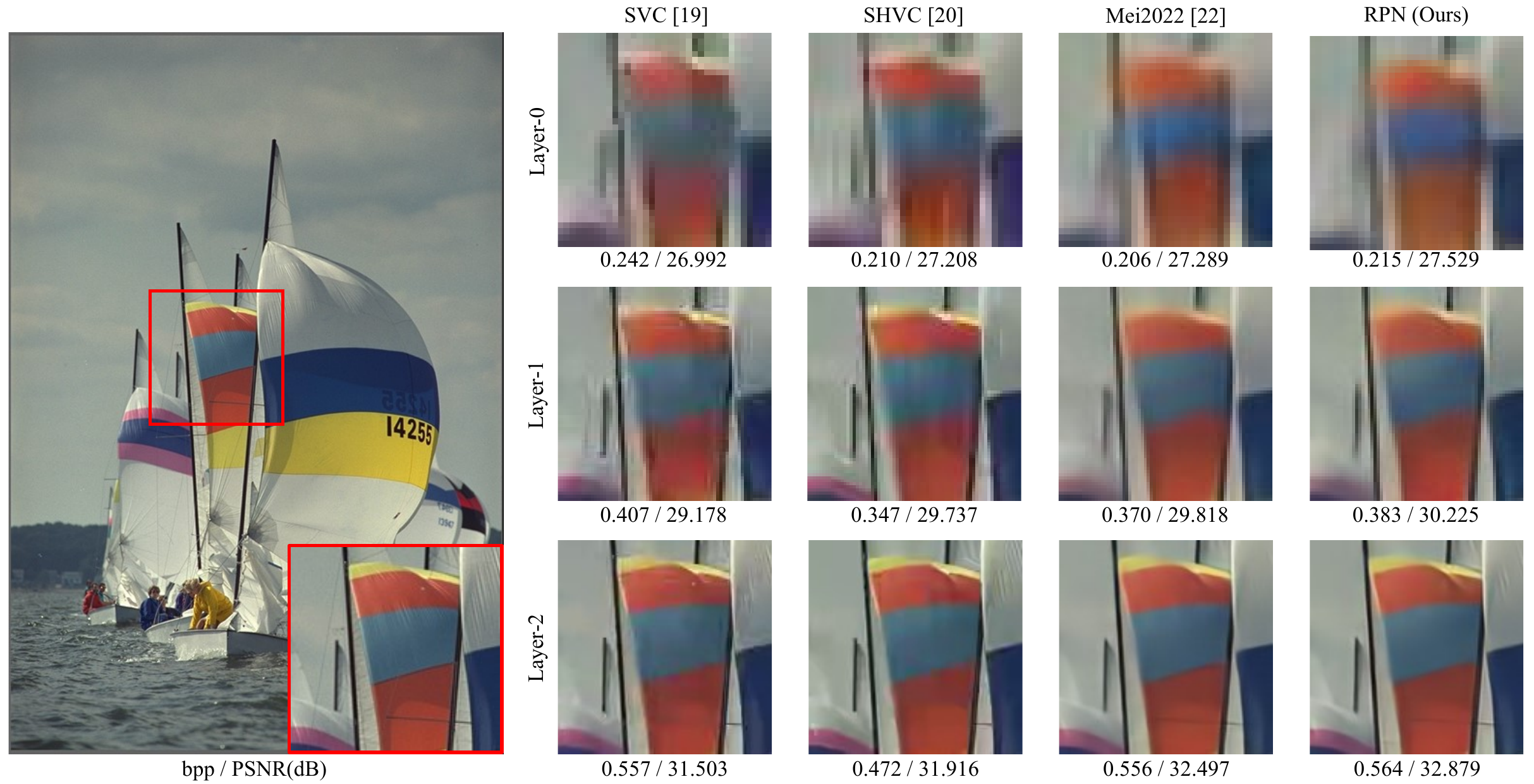}
\caption{Qualitative results with existing spatial scalable algorithms at each certain spatial resolution.} 
\label{fig7}

\end{figure*}

\begin{figure*}[t]

\centering
\setlength{\abovecaptionskip}{-0.2cm}
\includegraphics[width=1\linewidth]{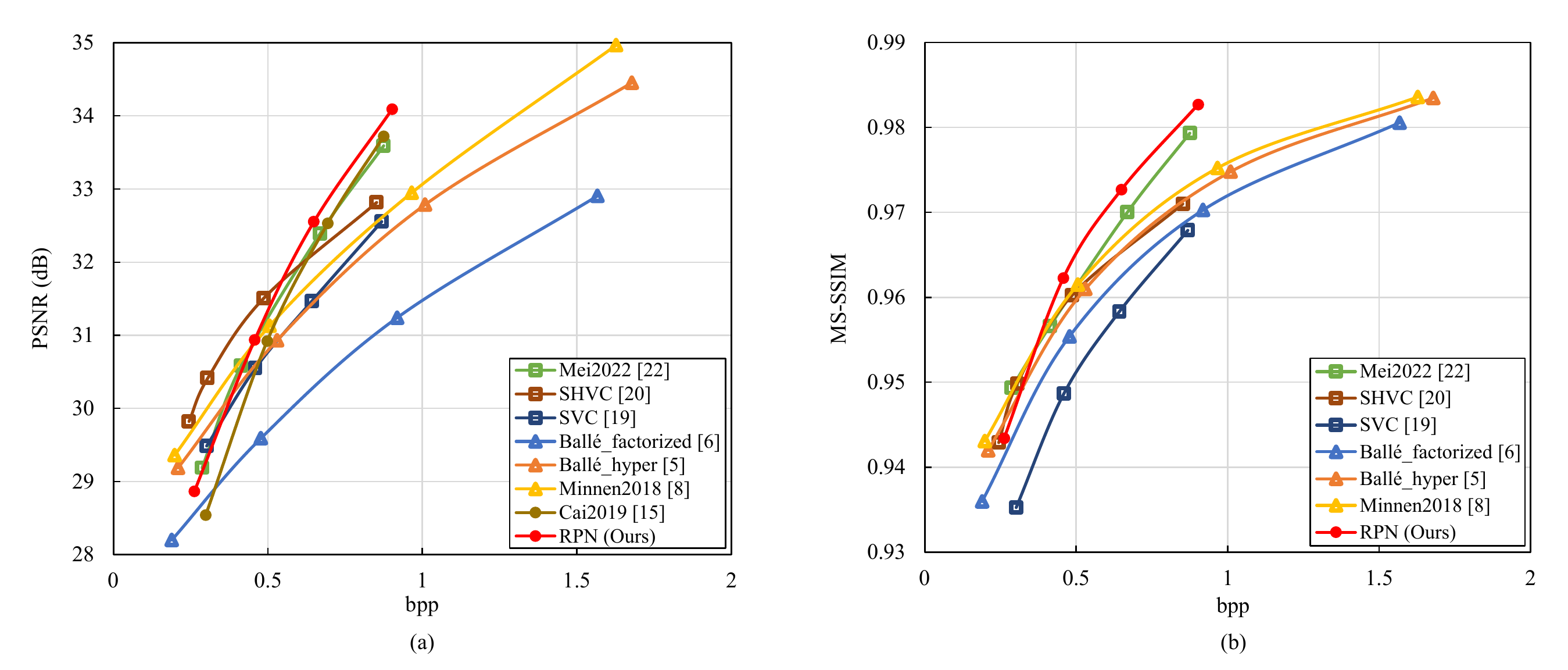}
\caption{Rate-distortion performance evaluation with existing compression methods at quality scalability in terms of (a) PSNR and (b) MS-SSIM.}
\label{fig8}

\vspace{0.2cm} 

\centering
\includegraphics[width=1\linewidth]{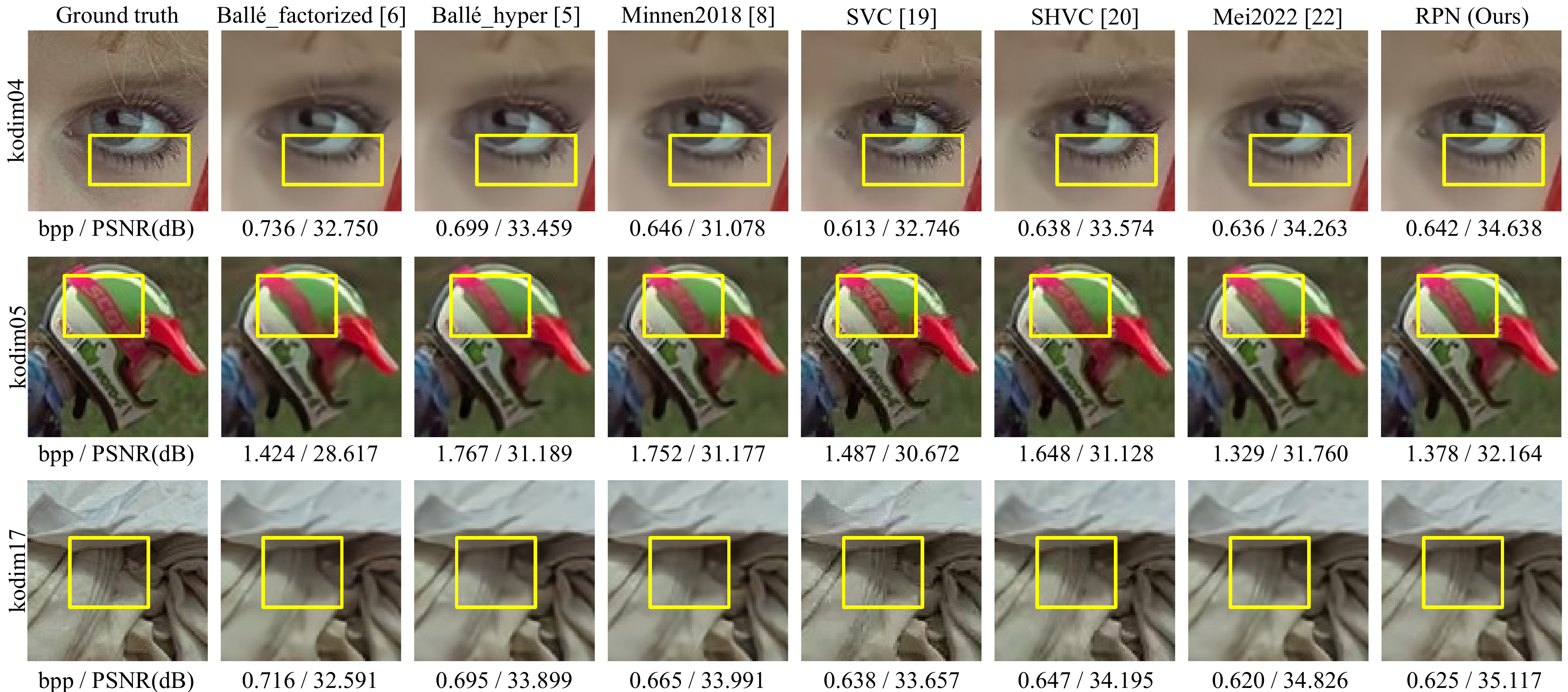}
\setlength{\abovecaptionskip}{-0.1cm}
\caption{Qualitative results of different methods under quality scalability.}
\label{fig9}

\end{figure*}

\begin{figure*}[t]
\centering
\includegraphics[width=1\linewidth]{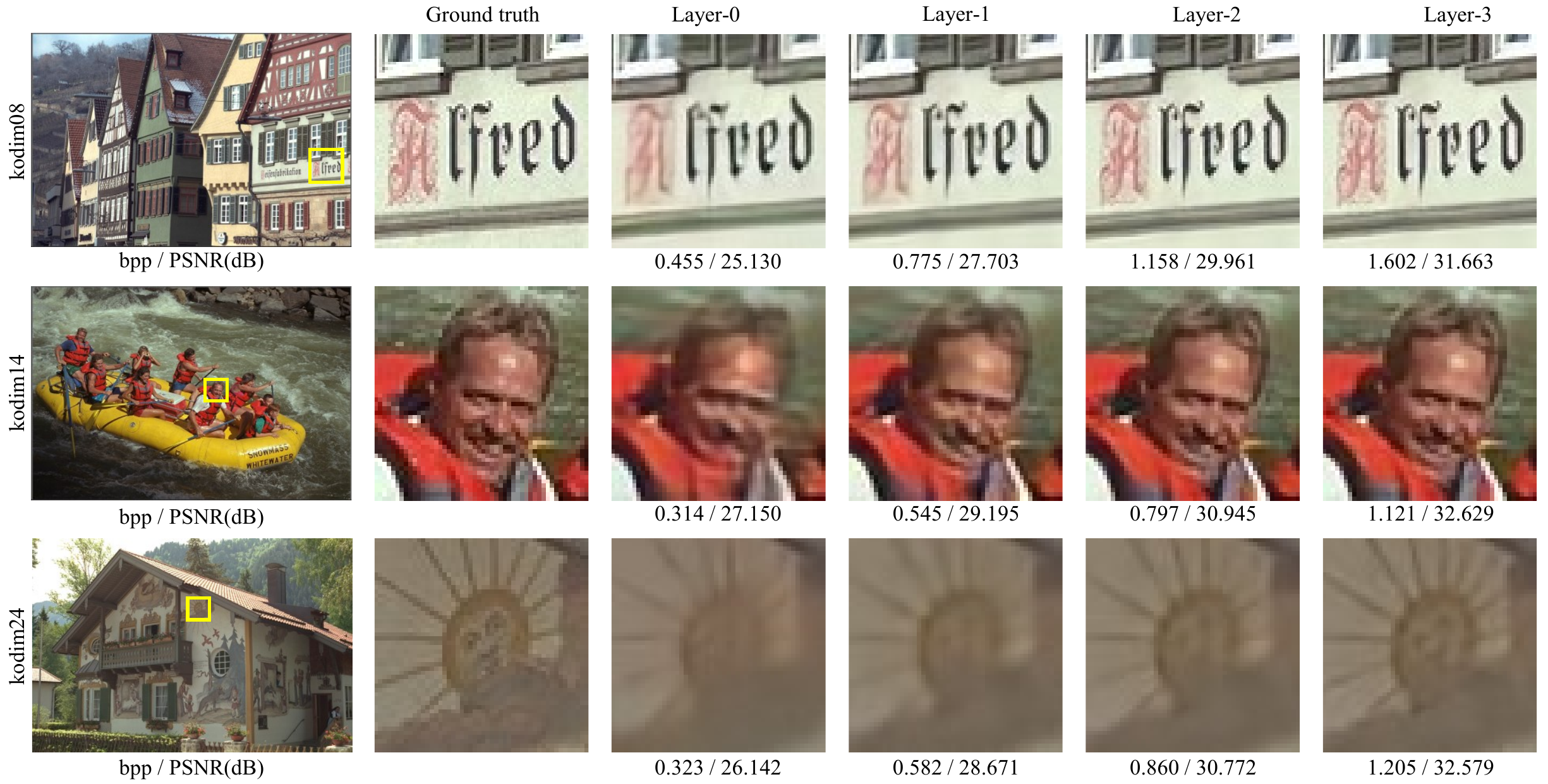}
\setlength{\abovecaptionskip}{-0.1cm}
\caption{Visual examples of decoded images of proposed RPN with quality scalability.}
\label{fig10}

\end{figure*}

\subsection{Comparison with State-of-the-Arts}
We compare the proposed RPN to existing state-of-the-art image compression codecs from both spatial and quality scalability perspectives.

\textbf{Spatial Scalable Image Compression}. We construct a 3-level RPN composed of one base layer and two enhance layers, which is consistent with the design methodology of \cite{mei2021learning}. We adopt two classical scalable compression codecs SVC~\cite{schwarz2007overview} and SHVC~\cite{boyce2015overview} and another four end-to-end learning-based image compression algorithms including: Ball{\'e}\_factorized~\cite{balle2017end}, Ball{\'e}\_hyper~\cite{balle2018variational}, Minnen2018~\cite{minnen2018joint}, Cai2019~\cite{cai2018efficient}, and Mei2022~\cite{mei2021learning}. All the results are produced by implementing their original public source codes. Besides, we use the software SHM-12.4~\cite{shm124} and JSVM-9.19~\cite{jsvm919} to perform SVC and SHVC compression, respectively. Since both two codecs are specially designed for video, we use FFmpeg~\cite{ffmpegtomar2006converting} to
convert the RGB image to YUV420 format for encoding,
and then convert the decoded image back to
RGB format for comparison. Ball{\'e}\_factorized~\cite{balle2017end}, Ball{\'e}\_hyper~\cite{balle2018variational}, Minnen \emph{et al.}~\cite{minnen2018joint} are all single-layer compression frameworks which have the same analysis transform and synthesis transform architectures as our RPN (see Table~\ref{tab1}). For fair comparisons, we utilize them to generate three independent bitstreams from $4\times\downarrow$ resolution to original resolution (\emph{i.e.} 4-2-1) to realize scalable coding, where the bitrates are denoted as $R_1$, $R_2$, and $R_3$. Therefore, according to Eq.~(\ref{eq1}), the bitrates of the base and two enhance layers are represented as $R_1$, $R_1+R_2$, and $R_1+R_2+R_3$, respectively. To our knowledge, Mei2022~\cite{mei2021learning} is the latest learning-based method for both spatial and quality scalable compression.

The R-D curves of all methods are illustrated in Fig.~\ref{fig5}. From Fig.~\ref{fig5}(a), we can see Mei2022~\cite{mei2021learning} exceeds SVC on all bpp in terms of PSNR at shows closely similar R-D performance to SHVC. Compared to these three scalable methods, our RPN achieves the best PSNR at all the base and enhance layers, which demonstrates its superiority of different spatial resolution versions compression. As for another three single-layer methods, \emph{i.e.} Ball{\'e}\_factorized~\cite{balle2017end}, Ball{\'e}\_hyper~\cite{balle2018variational}, Minnen2018~\cite{minnen2018joint}, it can be observed they perform better than other scalable methods and ours at the base layer but worst at the later two enhance layers. This is because these methods train and optimize each specific point on the R-D curve individually to pursue extreme performance. They ignore the gain effect of previous bitstreams in the current compression process. In contrast, RPN employs the previous bitstream to participate in current compression, which has good scalability and provides more optimal results with the resolution growing. This phenomenon can also be seen in Fig.~\ref{fig5}(b), these single-layer methods even experience performance degradation in terms of MS-SSIM after the base layer compression. The PSNR and MS-SSIM comparisons in Fig.~\ref{fig5} demonstrate the effectiveness of the proposed method for spatial scalable compression.

Fig.~\ref{fig6} shows the qualitative results of the final enhancement layer obtained by different methods. It can be seen that RPN can achieve efficient compression while preserving clearer textures, edges, and fewer blurs than other methods. To further investigate the reconstruction quality at each certain spatial resolution, we visualize the immediate results in Fig~\ref{fig7} and compare it to existing spatial scalable algorithms. As we can see, our RPN can produce multiple-resolution versions and show better local details at each spatial resolution.

\textbf{Quality Scalable Image Compression}. Following Mei2022~\cite{mei2021learning}, we use the same strategy to modify RPN, where all the downsampled images are replaced by the original one, constituting a 4-level quality scalable compression framework. Here, RPN is compared with SVC~\cite{schwarz2007overview}, SHVC~\cite{boyce2015overview}, Ball{\'e}\_factorized~\cite{balle2017end}, Ball{\'e}\_hyper~\cite{balle2018variational}, Minnen2018~\cite{minnen2018joint}, Cai2019~\cite{cai2018efficient} and Mei2022~\cite{mei2021learning} to validate its reconstruction fidelity. Fig.~\ref{fig8} illustrates the PSNR and MS-SSIM comparisons of all four compression levels from the base layers to the later three enhance layers.

As shown in Fig.~\ref{fig8}, though the typical single-layer methods Ball{\'e}\_hyper~\cite{balle2018variational} and Minnen2018~\cite{minnen2018joint} outperform all the scalable methods at the base layer and the first enhance layer, due to the scalability mechanism, they are surpassed at the last two enhance layer. Compared with scalable methods, from the first two R-D points in Fig.~\ref{fig8}(a), we can see the traditional codec SHVC~\cite{boyce2015overview} only shows higher PSNR than the learning-based methods Cai2019~\cite{cai2018efficient} and Mei2022~\cite{mei2021learning}. Our RPN shows better PSNR performance with less bpp over all these methods when coding bitstreams at later two enhancement layers. Especially, RPN performs better than Cai2019~\cite{cai2018efficient} and Mei2022~\cite{mei2021learning} at all R-D points. In Fig.~\ref{fig8}(b), it can be observed that the scalable methods show more obvious superiority than single-layer methods in terms of MS-SSIM. RPN achieves the best R-D performance at almost all points, which demonstrates that RPN can provide a more efficient and effective framework for quality scalable image compression.

\begin{table}[t]
  \centering
  \caption{PSNR/MS-SSIM \emph{vs.} Bit-rate Results using the Bjøntegaard Delta (BD) metric~\cite{bdbr} with SVC~\cite{schwarz2007overview} as the Anchor. \textbf{Text} Indicates the Best Performance.}
    \begin{tabular}{c|c|c}
    \hline
    \hline
    Methods & BD-rate (PSNR) & BD-rate (MS-SSIM)  \\
    \hline
    SHVC~\cite{boyce2015overview}  & \textbf{-25.38\%}  &  -31.35\% \\
    Ball{\'e}\_factorized~\cite{balle2017end} & 55.38\% & -21.29\% \\
    Ball{\'e}\_hypre~\cite{balle2018variational}  & -1.47\% & -29.74\%\\
    Minnen2018~\cite{minnen2018joint} & -13.63\% & -34.64\%\\
    Cai2019~\cite{cai2018efficient} & -5.37\% & --\\
    Mei2022~\cite{mei2021learning} & -13.30\% & -31.69\% \\
    RPN (Ours) & -16.83\% & \textbf{-37.10\%} \\
    \hline
    \hline
    \end{tabular}%
  \label{tab3}%
\end{table}

\begin{table*}[t]
\renewcommand\arraystretch{1.3}
  \centering
  \caption{The Model Complexity (Network Parameters) and Averaged Inference Speed (Second) Comparison On Kodak Dataset.}
    \begin{tabular}{c|c|c|c|c|c|c}
    \hline
    \hline
    \multirow{2}{*}{Method} & \multicolumn{3}{c|}{Spatial Scalable} & \multicolumn{3}{c}{Quality Scalable} \\
\cline{2-7}             & Parameters & Encoding time & Decoding time & Parameters & Encoding time & Decoding time \\
    \hline
Ball{\'e}\_factorized~\cite{balle2017end} & 5.7M  & 0.08  &  0.93  & 5.7M & 0.11 & 0.99 \\
Ball{\'e}\_hypre~\cite{balle2018variational} & 13.1M  & 0.13  &  1.01  & 13.1M & 0.16 & 1.04 \\
Minnen2018~\cite{minnen2018joint}  & 14.5M  & \textbf{0.62}  &  \textbf{2.41}  & 14.5M & \textbf{0.70}  & \textbf{2.65} \\
Mei2022~\cite{mei2021learning} & \textbf{32.0M}       & 0.10       & 0.34       & \textbf{99.7M}       & 0.13       & 0.37 \\
    RPN (Ours) & 18.1M       & 0.21       & 0.38       & 62.6M       & 0.23       & 0.47 \\
    \hline
    \hline
    \end{tabular}%
  \label{tab_compare}%
\end{table*}%

\begin{figure*}[t]
\centering
\setlength{\abovecaptionskip}{-0.1cm}
\includegraphics[width=1\linewidth]{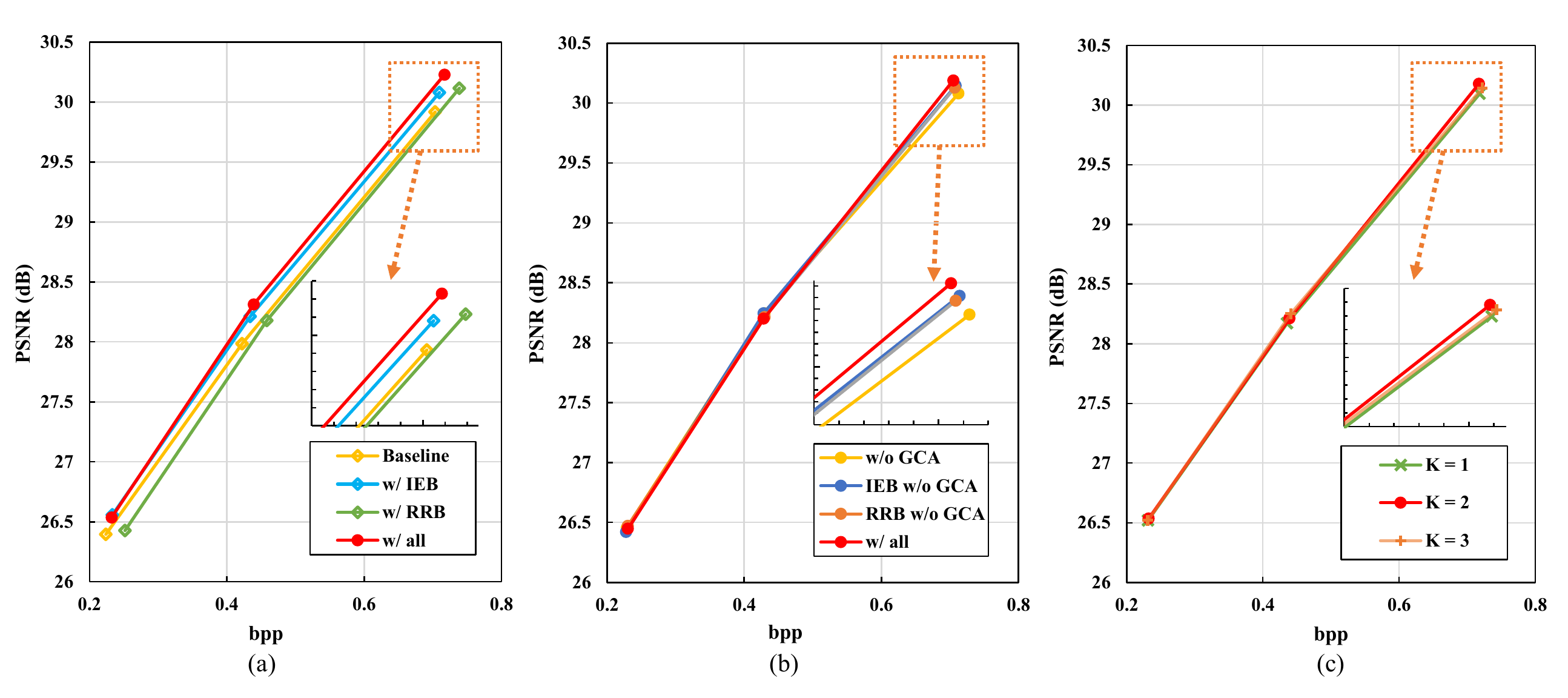}
\caption{Ablation study results of CRCM with (a) IEB and RRB, (b) GCA, and (c) the iteration number $K$.}
\label{fig11}
\end{figure*}

\textbf{BD-rate Saving Comparison}. Besides the R-D curves, we also choose SVC as the anchor and summarize the PSNR/MS-SSIM \emph{vs.} bitrate results using Bjøntegaard Delta (BD) metric~\cite{bdbr} on the Kodak dataset to measure the BD-rate savings between ours and compared methods. As shown in Table~\ref{tab3}, in terms of PSNR, Ball{\'e}\_factorized~\cite{balle2017end} surprisingly involves the largest bitrate cost than other methods. Compared with Ball{\'e}\_hypre~\cite{balle2018variational} and Cai2019~\cite{cai2018efficient}, RPN shows noticeable bitrate savings and also exceeds the other two methods Minnen2018~\cite{minnen2018joint} and Mei2022~\cite{mei2021learning} by a considerable margin. In combination with the results in Fig.~\ref{fig8}, SHVC can save more bitrate than RPN but exhibits obviously lower PSNR at higher bitrate. As for MS-SSIM, RPN achieves 37.10\% bitrate savings relative to SVC, which outperforms all the methods even including SHVC. These comparisons demonstrate that RPN can achieve a good trade-off between compression quality and efficiency.

In Fig.~\ref{fig9}, we provide several compressed images by different methods for qualitative comparison. The results indicate that RPN can reconstruct the images with better visual-pleasant quality, higher PSNR, and relatively lower bpp. Besides, an example of the reconstructed images from each layer in RPN is sketched in Fig.~\ref{fig10}, it can be seen that RPN can gradually improve the reconstruction sharpness and accuracy with the continuous compression layers following our quality scalable scheme.

\textbf{Model Complexity}. Furthermore, we compare the model complexity and efficiency with several methods on the Kodak dataset, which are measured by network parameters and inference speed respectively. Here, we calculate the encoding time and decoding time to comprehensively evaluate the model efficiency. For fair comparison, all the models are implemented on the same GPU using their public codes. Since Ball{\'e}\_factorized~\cite{balle2017end}, Ball{\'e}\_hypre~\cite{balle2018variational}, and Minnen2018~\cite{minnen2018joint} are designed for specific compression rates rather than scalable compression, we count the total running time for multiple bitrates which are consistent to Mei~\emph{et al.}~\cite{mei2021learning} and our models. 
As shown in Table~\ref{tab_compare}, it is natural that Ball{\'e}\_factorized~\cite{balle2017end} and Ball{\'e}\_hypre~\cite{balle2018variational} with single-layer framework involve fewer parameters and less encoding time cost.
Due to the incorporation of the autoregressive model, the codec of Minnen2018~\cite{minnen2018joint} has fewer parameters but the longest encoding time.
However, we can see that these three methods have higher decoding times than other methods.
Though Mei~\emph{et al.} shows better efficiency, it suffers from heavy model complexity. 
In contrast, our cross-resolution context mining module only needs fewer channels to mine enough contextualized information, thus our RPN can save about 37\% parameters (62.6M v.s. 99.7M) but just sacrifice very slight speed (about 0.1 seconds).

\subsection{Ablation Study}

Here with the spatial scalable model as an example, we conduct the ablation study experiments.
First, we investigate the effectiveness of all the proposed components in RPN including: the cross-resolution context mining module (CRCM) composed of redundancy removal block (RRB), information enhancement block (IEB), and global context attention (GCA). Then, we analyze the influence of the uncertainty guidance strategy.

\subsubsection{Effectiveness of RRB} 
The goal of RRB is to remove redundant information in received features based on spatial and channel sparsity. To study the effect of our RRB in CRCM, 
we first train a model without IEB, \emph{i.e.} ``w/ RRB'', and then replace RRB with vanilla convolutions to obtain a baseline model.
 As illustrated in Fig.~\ref{fig11}(a), even though we use a sparse implementation, the model with RRB still achieves comparable R-D performance against the baseline.
This is because RRB removes a large amount of information while retaining important context, resulting in feature distillation.

\subsubsection{Effectiveness of IEB} 
We also use the baseline model as the reference and directly incorporate IEB into it to verify the influence.
In Fig.~\ref{fig11}(a), it can be observed the model with IEB, \emph{i.e.} ``w/ IEB'', performs better than the baseline. Moreover, with the combination of IEB and RRB, \emph{i.e.} ``w/ all'', the full model can achieve the best R-D performance. On one hand, IEB can enrich the contextualized information of incoming features. On the other hand, iterative IEBs and RRBs in CRCM perform alternative feature enrichment and distillation, realizing complementary to each other. Therefore, the full model can produce more informative representations, thus revealing its superiority.

\subsubsection{Effectiveness of GCA} In CRCM, the GCA is both used in RRB and IEB to aggregate the features of all positions and form a global feature with more contextualized information. First, we directly remove all the GCAs in these blocks. As shown in Fig.~\ref{fig11}(b), the model (``w/o GCA'') shows the worst performance. When we only retain the GCA in RRBs (``IEB w/o GCA'') or IEBs (``RRB w/o GCA''), the performance of these two models is obviously improved. By employing GCA in both RRBs and RRBs, it can be seen the full model (``w/ all'') achieves the best performance, which demonstrates its effectiveness for image compression.

\begin{figure}[t]
\centering
\includegraphics[width=1.0\linewidth]{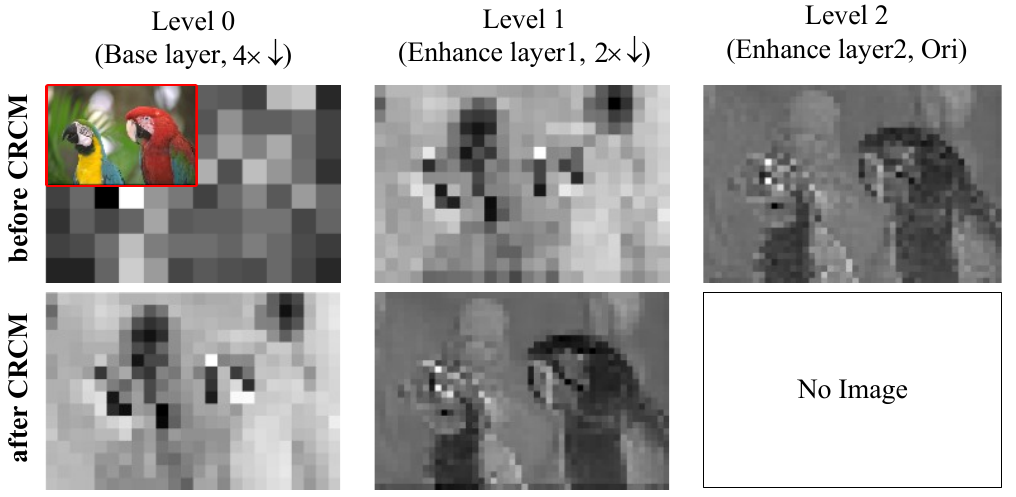}
\caption{The visualization of the feature maps of kodim23 before and after the CRCM. ``$4 \times \downarrow$'' and ``$2 \times \downarrow$'' represents the downsampling factor of feature resolution in level $0$ and level $1$ compared to the size in level $3$ (Ori).}
\label{fig12}
\end{figure}

\subsubsection{Iteration Number of IEB and RRB} In CRCM, the IEB and RRB are iteratively implemented for alternative mining cross-scale context in features. Here, we investigate the influence of the total iteration number $K$ for image compression, where $K=1$ denotes the ``IEB-RRB-IEB'' manner in one CRCM. As shown in Fig.~\ref{fig11}(c), by comparing $K=1$ and $K=2$, we can see larger $K$ obtain a slight advantage in PSNR score at the same bpp. When we increase $K$ to 3, it can be observed that although the model involves larger complexity, its performance has reached saturation or even decreased. Therefore, in this work, we adopt $K=2$ as the final setting for the trade-off between R-D performance and model complexity.

\subsubsection{Effectiveness of the Overall CRCM} To intuitively understand the resolution fields exploration of CRCM, we visualize the average feature maps before and after the CRCM in each compression level. As shown in Fig.~\ref{fig12}, we use a 3-level model (4-2-1) for validation. Note that all the maps are zoomed in to the same resolution for convenient comparison. It can be observed that the feature after CRCM contains clearer structures and richer details than its input one. Between two adjacent levels, this phenomenon indicates that such a resolution field can provide informative contexts to help the following compression. By using it as the residual prior, we can minimize the bitstream required in the next compression level while preserving fine-grained textures.

\subsubsection{Influence of Uncertainty Guidance Strategy} 
\begin{figure}[t]
\centering
\includegraphics[width=1.0\linewidth]{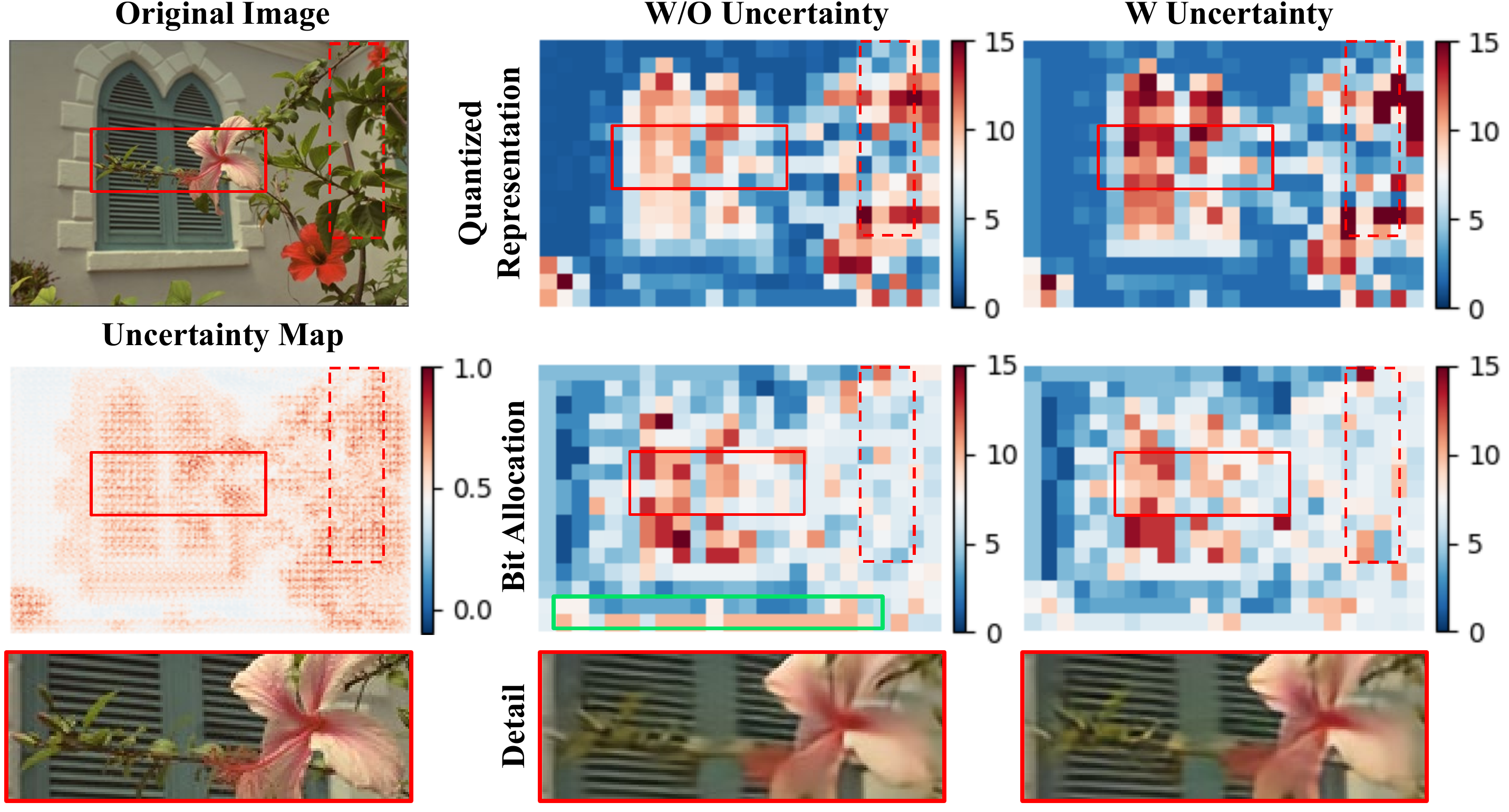}
\caption{Visualization of the estimated uncertainty map of Level 1, the detail of the decoded image, the intermediate quantized representation, and the bit allocation of the model with and without uncertainty-guided optimization.}
\label{fig_13}
\end{figure}

In RPN, we construct a reverse bottom-up pyramid process that progressively conducts explicitly aleatoric uncertainty estimation from high-to-low resolutions to make a more reliable and accurate reconstruction in low compression levels. As described in Section~\ref{strategy}, our uncertainty guidance strategy is reflected by the designed uncertainty-guided loss, which contains three training steps. 
To fully investigate its influence, in Figure~\ref{fig_13}, we visualize the estimated uncertainty map of Level 1, the intermediate quantized representation and the bit allocation of the model with and without uncertainty-guided optimization. 
Besides, we also give the detail part of the decoded image.
As we can see, the uncertainty map reveals the local region ({\color{red}{red rectangle}}) with abundant textures or structural edges that should be emphasized, which are important for accurate image reconstruction. It can also be observed that the model with uncertainty guidance can force networks to focus on the important regions corresponding to the uncertainty map and make a more reasonable bit allocation. In contrast, the vanilla model without uncertainty guidance even allocates considerable bits on less informative regions ({\color{green}{green rectangle}}). 

\begin{table}[t]
\renewcommand\arraystretch{1.2}
  \centering
  \caption{Ablation study of the uncertainty-guided loss and training strategy.}
    \begin{tabular}{@{}c@{}|c|c|@{}c@{}}
    \hline
    \hline
    \multirow{3}{*}{\makecell{Compression \\ Levels}} & \multirow{3}{*}{Metrics} & w/o uncertainty & w/ uncertainty \\
    \cline{3-4}
    ~ & ~ & Step 1 & Step  2 \& Step 3\\
    \cline{3-4}
    ~ & ~ & $\mathcal{L}_{sca}$ & $\mathcal{L}_{U}+\mathcal{L}_{sca}$ \& $\mathcal{L}_{UG}+\mathcal{L}_{sca}$ \\

    \hline
    \multirow{3}[0]{*}{Level 0} & bpp     & 0.2325  & \textbf{0.2245} \\
\cline{2-4} & PSNR    & 26.5358 & \textbf{26.5478} \\
\cline{2-4}  & MS-SSIM    & 0.9296  & \textbf{0.9314} \\
    \hline
    \multirow{3}[0]{*}{Level 1} & bpp     & 0.4393  & \textbf{0.4303} \\
\cline{2-4} & PSNR    & 28.3127 & \textbf{28.3267} \\
\cline{2-4} & MS-SSIM    & 0.9404  & \textbf{0.9419} \\
    \hline
    \multirow{3}[0]{*}{Level 2} & bpp     & 0.7165  & \textbf{0.7152} \\
\cline{2-4} & PSNR    & 30.1569 & \textbf{30.189} \\
\cline{2-4} & MS-SSIM    & 0.9513  & \textbf{0.9549} \\
    \hline
    \hline
    \end{tabular}%
  \label{tab2}%
\end{table}%

In addition, we report the quantitative performance of these two models in terms of bpp, PSNR, and MS-SSIM in Table~\ref{tab2}. Firstly, we can observe that, equipped with the uncertainty-guided loss, the network can achieve higher PSNR and MS-SSIM with smaller bpp. 
Then, it should be noted that our uncertainty guidance scheme aims at guiding the network to emphasize the textures and edges that have high variance (uncertainty) as illustrated in Fig.~\ref{fig_13}. As shown in the detail parts, the flower and leaves with uncertainty-guided optimization have more reliable texture reconstruction.  

\section{Conclusion}

In this paper, we explore the potential of the resolution field and aleatoric uncertainty for scalable image compression, thus presenting the reciprocal pyramid network (RPN). RPN is a multi-level compression framework that consists of a forward top-down compression pyramid and a reverse bottom-up uncertainty-guided optimization pyramid to achieve both spatial and quality scalability. In the forward pyramid, between adjacent compression levels, we propose a cross-resolution context mining module (CRCM) to generate the informative resolution field as residual prior to help the current compression level. In the reverse pyramid, we characterize the textured pixels with large variance as the uncertainty parts and conduct uncertainty-guided optimization from high-resolution space to low-resolution space. Extensive experimental results demonstrated the superiority of our RPN against existing scalable image compression methods on both R-D performance and visual quality.

From the results in Table~\ref{tab3}, the compression efficiency still needs to be further improved. Therefore, in our future work, we will investigate a more efficient yet effective scalable compression framework to achieve an optimal trade-off between R-D performance, model complexity, and inference speed. Secondly, we will combine scalable compression with downstream tasks to study a new paradigm that balances human and machine visions. Lastly, we will also conduct more research on context modeling, such as powerful architectures and entropy models, for better compression and reconstruction.


%






\ifCLASSOPTIONcaptionsoff
  \newpage
\fi



\bibliographystyle{IEEEtran}
\bibliography{IEEEabrv, egbib_my}
\end{document}